# Time Varying Risk Aversion: An Application to Energy Hedging*

## John Cotter[a] and Jim Hanly[b]

August 2009

[a]John Cotter,
Director of Centre for Financial Markets,
School of Business
University College Dublin,
Blackrock,
Co. Dublin,
Ireland,
Tel +353 1 716 8900,
e-mail john.cotter@ucd.ie.
Corresponding Author

[b]Jim Hanly,
School of Accounting and Finance,
Dublin Institute of Technology,
Dublin 2,
Ireland.
tel +353 1 402 3180,
e-mail james.hanly@dit.ie.

*The authors would like to thank participants at a University College Dublin seminar for their comments on an earlier draft. Part of this study was carried out while Cotter was visiting the UCLA Anderson School of Management and is thankful for their hospitality. Cotter's contribution to the study has been supported by a University College Dublin School of Business research grant. The authors thank two anonymous referees for very helpful comments but the usual caveat applies.

# Time Varying Risk Aversion: An Application to Energy Hedging

## Abstract

Risk aversion is a key element of utility maximizing hedge strategies; however, it has typically been assigned an arbitrary value in the literature. This paper instead applies a GARCH-in-Mean (GARCH-M) model to estimate a time-varying measure of risk aversion that is based on the observed risk preferences of energy hedging market participants. The resulting estimates are applied to derive explicit risk aversion based optimal hedge strategies for both short and long hedgers. Out-of-sample results are also presented based on a unique approach that allows us to forecast risk aversion, thereby estimating hedge strategies that address the potential future needs of energy hedgers. We find that the risk aversion based hedges differ significantly from simpler OLS hedges. When implemented in-sample, risk aversion hedges for short hedgers outperform the OLS hedge ratio in a utility based comparison.

## 1. Introduction

Offsetting energy price risk is becoming increasingly important given the recent volatility in world energy markets and the strong links between energy prices and macroeconomic activity (Sadorsky, 2006). In this context the level of risk aversion of hedging market participants is a key input in the estimation of hedging strategies based on Expected-Utility Maximisation (EU). A risk aversion measure that is commonly used in this framework is the Coefficient of Relative Risk Aversion (CRRA). However, in the hedging literature, no study [1] has explicitly calculated the level of risk aversion of hedgers, despite its key role in the estimation of optimal hedge strategies. Instead, a range of arbitrary values have been applied that reflect general risk averse preferences [2]. This paper addresses this issue, using a novel approach that estimates and applies the observed risk aversion of energy hedgers, to generate utility maximizing hedge strategies based on the unleaded gasoline market. Given that the risk aversion parameter has a large influence on the hedge ratio (HR), the use of arbitrary measures will yield strategies that do not reflect hedger's actual attitudes to risk and may result in suboptimal hedging solutions.

An additional shortcoming of the literature is that it has tended to focus on the effect that different risk preferences will have on the hedge strategy but not on the performance of the hedging strategy. This is an important issue, as although hedging has a benefit through risk reduction, it also has a cost in terms of the potential loss of expected return (Kahl, 1983). Finally, while the literature has acknowledged the importance of allowing the hedge ratio to vary over time (Cecchetti et al, 1988), no provision has been made to incorporate time-varying risk aversion in the estimation of optimal hedges, despite evidence that risk aversion is time-varying (see, for example, Campbell and Cochrane, 1999).

---

[1] Values of the CRRA have been calculated to match the data in various economic and financial models but not, to our knowledge, in the hedging literature. See, for example, Mehra and Prescott (1985) who estimated that a large value of the CRRA (in excess of 10) was required in order to reconcile the large premium earned by equity returns with the return on risk free securities.

[2] For example, Kroner and Sultan (1993) use values of 4 and 6 for the risk aversion parameter by taking these estimates from the equity pricing literature. The Lower Partial Moment framework also includes risk aversion but the values are arbitrarily chosen rather than being related to actual attitudes to risk.

This paper makes a number of contributions to address these issues. Firstly, we estimate the risk aversion of energy market participants and apply it to derive a time varying hedge ratio that explicitly incorporates risk aversion. Rather than applying arbitrary values for the risk aversion parameter, we base our estimates on the risk preferences of unleaded gasoline market participants. Secondly, we explicitly estimate a time-varying market representative risk aversion coefficient using a GARCH-M model. This is calculated separately for both long and short hedgers and for weekly and monthly data, and we compare these for both hedgers and across frequencies. Thirdly, we apply the resulting observed risk preferences in the estimation of a time-varying optimal hedging strategy, where optimal is defined as the utility maximizing hedge for a given level of risk aversion. In this way, hedging solutions reflect the level of risk aversion in the market, and account for both risk and expected return. It also allows us to compare the hedging outcomes of both short and long hedgers with those from the EU literature. Also, we compare the risk aversion hedge strategies with some commonly applied hedge strategies, using a measure of hedging effectiveness based on utility maximisation[3]. Finally, in this paper we examine forecasted risk aversion to estimate utility maximizing hedge strategies that address the potential future needs of energy hedgers.

Empirical results indicate that there are significant differences between the utility maximising and the variance minimising hedges. We find that in-sample, the risk aversion hedges outperform the simpler OLS model for the short hedgers. However, out-of-sample, the results tend to favour the OLS model. We also find significant differences between the risk aversion of short as compared with long hedgers. These risk preferences differ considerably from the arbitrary values applied in the EU hedging literature.

The remainder of this paper proceeds as follows. Section 2 details the optimal hedging framework and the role of risk aversion in the estimation of optimal hedging strategies. In Section 3 we define the CRRA and outline the framework used in estimating it, together with a brief review of the risk aversion literature. Section 4 describes the data and estimation methodology. Empirical results are presented in section 5 and concluding remarks in section 6.

## 2. The Hedging Problem

---

[3] We also examine typically used hedge strategies including an un-hedged position (no-hedge), and the variance minimizing hedge.

In determining what is optimal in a hedge strategy, we can distinguish between two different approaches. In the hedging literature, a return volatility framework based on variance minimisation (see, for example, Johnson, 1960) has become dominant. This approach equates optimal with risk reduction, therefore strategies that choose a hedge ratio to minimise risk are referred to as the minimum variance hedge ratio (MVHR). The advantages of this framework are its ease of calculation and interpretation. However, the risk minimizing framework focuses purely on risk, and doesn't explicitly consider either the level of risk aversion, or the expected return in the design of an optimal hedging strategy. Also, the variance based literature does not differentiate between short and long hedgers and thus ignores the idea that different types of hedgers may have different attitudes towards risk.

The second approach to hedging is to maximize expected utility, where utility is a function of both risk and expected return. This framework incorporates risk aversion as a key element in the estimation of the hedge strategy. Many papers have avoided making a distinction between the utility maximizing and the variance minimising approaches by assuming that futures prices follow a martingale (Lence 1995). Under this assumption, the MVHR and the utility maximizing hedge ratios are equivalent, however, there is evidence (Moosa and Al-Loughani, 1994) that Oil futures markets do not follow a martingale. In this paper, we focus our attention on the estimation of hedge strategies that maximize expected utility. This allows us to incorporate time-varying risk aversion within the hedging context and to estimate hedges that are based on both risk and expected return. These hedges are also tailored to the individual risk preferences of hedgers. We term hedges that are estimated in this way - the risk aversion hedge ratio (RAHR). We also estimate the MVHR and compare the two approaches to see which dominates in terms of expected utility. First we briefly define two different types of hedgers.

**Definition of Hedgers**

For most of the literature, no distinction is drawn between short and long hedgers despite the fact that they may have widely differing reasons for participating in the futures market. In this paper, we distinguish two sets of hedgers; short hedgers and long hedgers. Within a commodity hedging setting, a short hedger may be regarded as a producer (e.g. oil companies) and so is concerned with price decreases, whereas the long hedger is typically a user of refined oil products and so will be concerned with price increases. Therefore, they are interested in opposite sides of the return distribution. Using our example of hedging energy price risk, the long hedgers would be large users of refined oil products and short hedgers are the oil companies. Short and long hedgers in oil markets have different characteristics. For example the evidence[4] on oil futures market participation has shown that long hedgers (consumers) tend to be more active than short hedgers (producers). Given their different participation in the futures market, it would be reasonable to expect them to have different levels of risk aversion and associated hedging strategies. Therefore, they should be considered separately.

---

[4] See for example Devlin and Titman (2004) who find that oil producers tend to hedge only the difference between their current production and the minimum economic production level. Also, the high correlation between oil company profits and the oil price suggests that oil producers do not hedge a large proportion of their production.

**Optimal Hedge Ratio's under the EU Framework**

An optimal hedging strategy for both short and long hedgers can be derived as follows. Assuming a fixed spot position let $r_{st}$ and $r_{ft}$ be logarithmic returns on the spot and futures series respectively, and $\beta$ be the Optimal Hedge Ratio (OHR). The return to the hedged portfolio is constructed as follows:

$$R_p = +r_s - \beta r_f \qquad \text{(short hedger)} \qquad (1a)$$

$$R_p = -r_s + \beta r_f \qquad \text{(long hedger)} \qquad (1b)$$

The short hedger is long the spot asset and is concerned with negative returns. For the long hedger the position is reversed. The OHR is the weight of the futures asset in the hedged portfolio that is chosen either to minimise risk or to maximize expected utility, depending on the underlying framework that is being applied. Assuming that the agent has a quadratic utility function[5], then the OHR can be calculated as:

$$\beta = \frac{E(r_{ft})}{2\lambda \sigma_{ft}^2} + \frac{\sigma_{sft}}{\sigma_{ft}^2} \qquad (2)$$

where $E(r_{ft})$ is the expected return on futures, $\lambda$ is the risk aversion parameter, $\sigma_{ft}^2$ is the futures variance and $\sigma_{sft}$ is the covariance between spot and futures. Equation (2) thus explicitly establishes the relationship between the risk aversion parameter $\lambda$ and the OHR. The first term contains the speculative element of the hedge strategy, the second term is the hedging or risk minimising element[6]. As risk aversion increases, the individual hedges more and speculates less relative to the spot position, such that for extremely large levels of risk aversion, the first term will approach zero. Therefore, the OHR under the variance minimising framework given by (2) will become

$$\beta = \frac{\sigma_{sft}}{\sigma_{ft}^2} \qquad (3)$$

---

[5] Levy and Markowitz (1979) show that maximizing the mean-variance objective function provides a good approximation of maximizing expected utility regardless of the distribution of returns or the utility function chosen.

Similarly, under the assumption that the futures price follows a martingale[7], $E(r_{ft})=0$, (2) reduces to (3). Thus, under the assumptions of either infinite risk aversion, or that futures prices are unbiased, the utility maximizing hedge and the variance minimising hedge are equivalent. Given the role played by the expected return on futures $E(r_{ft})$ in the utility maximizing hedge, hedge ratios will be driven by the speculative element for lower levels of risk aversion. The assumption of infinite risk aversion is clearly incorrect given the wealth of evidence for a finite risk aversion value (Ait-Sahalia and Lo, 2000). There is also evidence that oil futures prices may not follow a martingale (Sadorsky, 2002)[8]. Therefore, we use the utility maximizing framework which allows the risk aversion of an investor, to impact on the choice of the OHR through (2). Consequently, we draw a distinction between the variance minimising hedge and the utility maximizing hedge. For comparison purposes, we also calculate a hedge strategy for the minimum variance investor using (3).

---

[6] Duffie (1989) shows that the optimal hedge ratio for a person with mean-variance utility can be decomposed into two terms: one speculative (which varies across individuals according to their risk aversion) and another reflecting a pure risk reduction component
[7] Under the martingale assumption i.e. that the expected return on futures is zero, the expected returns from a hedged portfolio will be unaffected by the number of futures contracts held and therefore the risk minimizing hedge becomes equivalent to the utility maximizing hedge.
[8] Sadorsky's reasoning is based on market efficiency in that the existence of a risk premium indicates that futures markets are not unbiased and this can be interpreted as evidence that they do not follow a martingale.

**Risk Aversion in the Hedging Literature**

Very few papers have examined utility maximizing hedge strategies based on the variance risk measure. Brooks, Cerny and Miffre (2007) incorporate risk aversion in a utility based hedging framework and their findings indicate that utility based hedges outperform OLS hedges in-sample, however, this finding didn't persist out-of-sample where the OLS model tended to perform best. More recently, deVille deGoyet, Dhaene and Sercu, (2008) examine optimal hedges for a range of commodities using a framework that incorporates risk aversion. They find that risk aversion has a considerable effect on the optimal hedge resulting in OHR's as high as 2.37.

There are a number of shortcomings in terms of the application of risk aversion in the variance based hedging literature. In the first place, few papers have incorporated risk aversion despite its centrality to the idea of hedging. Secondly, of the few papers that have incorporated risk aversion, none of them allow the risk aversion parameter to vary over time. This is a key issue since the evidence clearly suggests that just as the covariance of asset price returns is conditionally time-varying so too is the relation between the expected risk premium and the variance (Campbell and Cochrane, 1999). A further problem is that only a small range of values have been applied. They reflect estimates of risk aversion taken from other literatures, and these values may not be appropriate for energy hedgers. The literature also tends to focus on the effects of differing levels of risk aversion on the hedge ratio, but little attention is paid to hedging performance relative to risk aversion for utility maximizing strategies.

## 3. Risk Preferences and the CRRA

The risk aversion of an investor is expressed by their utility function which reflects an investor's view of the tradeoff between risk and return. Absolute risk aversion (ARA) is a measure of investor reaction to dollar changes in wealth. We can measure this by the relative change in the slope function at a particular point in their utility curve[9]. The CRRA differs from the ARA in that it examines changes in the relative percentages invested in risky and risk free assets as wealth changes. We define it as follows:

$$\text{CRRA} = -W * \frac{U^{''}(Wealth)}{U^{'}(Wealth)} \qquad (4)$$

Thus, it is similar to absolute risk aversion but with a scaling factor to reflect the investors current level of wealth (Arrow, 1971). The CRRA represents the investor's attitude towards risk in a single number, and will materially impact the choice of HR as previously discussed. In this paper we view the CRRA within its role as a determinant of the market risk premium. This framework is outlined in the next section.

---

[9] This refers to assumptions re changes in risk preferences as wealth changes. To measure an investor's absolute risk aversion we use

$$\frac{-U^{''}(Wealth)}{U^{'}(Wealth)}$$

**Derivation of the CRRA**

Estimation of the CRRA is based on the market risk premium, defined as the excess return on a portfolio of assets that is required to compensate for systematic risk[10]. Within the asset pricing framework, the size of the risk premium of the market portfolio is determined by the aggregate risk aversion of investors and by the volatility of the market return as expressed by the variance.

$$E(r_m) - rf = \lambda \sigma_m^2 \qquad (5a)$$

$$\frac{E(r_m) - rf}{\sigma_m^2} = \lambda \text{ (CRRA)} \qquad (5b)$$

where $E(r_m) - rf$ is the excess return on the market (or risk premium), $\lambda$ is the coefficient of relative risk aversion (CRRA) and $\sigma_m^2$ is the variance of the return on the market. Intuitively, the CRRA depends on the size of the risk premium associated with a given investment. Consequently, the CRRA is the risk premium per unit risk (Merton, 1980).

We estimate the CRRA using a framework that was developed by Frankel (1982) and adapted by Giovannini and Jorion (1989). This framework is based on a utility

---

[10] Systematic risk refers to market risk or risk that cannot be diversified away. Therefore investors who hold the market portfolio expect to be compensated for this minimal level of risk.

maximizing investor whose utility function is defined over the conditional expectation and conditional variance of end-of-period wealth[11]:

$$\max U\left[E_t(W_{t+1}), \sigma_t^2(W_{t+1})\right] \quad (6)$$

where

$$E_t(W_{t+1}) = W_t x_t' E_t(R_{t+1}) + W_t(1 - x_t'1)R_t^f \quad (7)$$

$$\sigma_t^2(W_{t+1}) = W_t^2 x_t' E\Omega_{t+1} x_t \quad (8)$$

where $W_t$ represents investors wealth and $x_t$ is the vector of investment shares in risky assets whose rates of return have conditional means and covariances denoted by $E_t(R_{t+1})$ and $\Omega_{t+1}$ respectively. $R_t^f$ is the risk free rate and 1 is a unit vector. The first order condition implies the following relationship between asset shares and the conditional moments of the return distributions:

$$x_t = \frac{(E_t(R_{t+1}) - R_t^f)}{\lambda \Omega_{t+1}} \quad (9)$$

where $\lambda$ represents the CRRA which is assumed to be constant. Equation (9) can be solved to obtain equilibrium expected returns using:

$$E_t(R_{t+1}) - R_t^f = \lambda \Omega_{t+1} x_t \quad (10)$$

and since $E_t(R_{t+1})$ is equal to the actual return less a forecast error, we have:

---

[11] We recognise that variance is not the only valid measure of risk and that there are other utility frameworks (e.g. see Cotter and Hanly, 2006), however we based our approach on the mean variance hedging framework which assumes quadratic utility as it is the most commonly applied approach in the hedging literature.

$$R_{t+1} = R_t^f + \lambda \Omega_{t+1} x_t + \varepsilon_{t+1} \qquad (11)$$

where $\varepsilon_{t+1}$ is the unexpected return and is orthogonal under rational expectations to all variables in an economic agent's information set. This general return volatility framework can be adjusted to account for any portfolio of assets, (e.g. hedged portfolios in 1a and 1b). These hedged portfolios do not include a risk free asset which we have assumed to be zero for simplification, but consist of just two assets, the unleaded gasoline spot and futures. $\Omega_{t+1} x_t$ in (11) is the variance of the portfolio which is simply a weighted average of the variances of the assets comprising the portfolio. In the hedging scenario, this term is replaced by the variance of the hedged portfolio. The adjusted equation can be written as:

$$R_{pt} = \lambda \sigma_{pt}^2 + \varepsilon_t \qquad (12)$$

In the next section we outline the model that is used to estimate (12).

**CRRA Estimation**

The model that we use is a univariate specification of the Diagonal Vech model proposed by Bollerslev, Engle and Wooldridge (1988). This model imposes a symmetric response on the variance and has been extensively applied in various literatures to model the variance (see, for example, Cotter and Hanly, 2006). We employ a GARCH-M specification of the model (Engle et al, 1987) to estimate (12).[12] This model was developed to allow investors be rewarded for additional risk by way of a higher return, with the mean equation adjusted to take account of the conditional variance of returns. The GARCH-M specification was chosen for a number of reasons. Firstly it provides us with a simple way of estimating risk aversion that is not too computationally intensive and is widely applied (e.g. Glosten et al 1993). This is important given that we are estimating risk aversion repeatedly. Secondly, the GARCH-M model allows us to simultaneously estimate the conditional mean and variance. Thirdly, it is designed to account for ARCH effects which are present in the data.

---
[12] They originally used an ARCH-M specification however it is more usual to use a GARCH-M specification given the advantages of the GARCH model over the ARCH.

From (12), $\lambda \sigma_{pt}^2$ is the risk premium, and the parameter $\lambda$ is the CRRA [13]. The conditional mean and variance of the hedged portfolio are modelled as follows:

$$r_{pt} = \lambda \sigma_{pt}^2 + \varepsilon_t \tag{13}$$

$$[\varepsilon_t]\Omega_{t-1} \sim N(0, \sigma_{pt}^2) \tag{14}$$

$$\sigma_{pt}^2 = c + a\varepsilon_{t-1}^2 + b\sigma_{pt-1}^2 \tag{15}$$

With (13) estimated with an intercept and where $r_{pt}$ is the return on the hedged portfolio, $\varepsilon_t$ is the residual, $\sigma_{pt}^2$ denotes the variance of the hedged portfolio, $\lambda$ is the CRRA, and $\Omega_{t-1}$ is the information set at $t-1$.

---

[13] In our model we use the contemporaneous value of the conditional variance, though the lagged value may also be used.

**Risk Aversion Estimates in Other Literatures**

Despite the importance of risk aversion in the EU hedging framework, no studies, to our knowledge, have explicitly estimated the risk aversion of a typical hedging agent. Instead, arbitrary values are chosen to reflect the generally accepted levels of risk aversion that have been found in other literatures. There has been little written regarding the risk aversion of investors in energy products. For this reason we looked at the equity literature for estimates of risk aversion. More generally, the literature on risk aversion has developed around early work by Arrow (1971), who argued on theoretical grounds, that the CRRA should be around 1. Other studies have differed widely in their estimates of risk aversion. For example, Mehra and Prescott (1985) required that the CRRA be in excess of 10 in order to reconcile the equity risk premium with theoretical models. More recently, both Brandt and Wang (2003) and Ghysels et al (2005) have estimated the CRRA to be in the region of 1.5 – 2 on average, while Guo and Whitelaw (2006) estimated it as 4.93. Rather than arbitrarily using these values from the equity literature, we instead estimate risk aversion based on investors in energy products which may yield different risk preferences.

**4. Data and Estimation**

Hedging is about reducing uncertainty, and we focus on energy hedging given the recent large price rises and subsequent decline, and volatility associated with energy based commodities. The energy contract used is NYMEX New York Harbor (HU)

Unleaded Gasoline.[14] This was chosen as it is the largest of the refined oil products in terms of traded volume, and because in times of uncertainty about supply, the convenience yield associated with Unleaded Gasoline may give rise to a higher risk premium.[15] Our full sample runs from 19/02/1992 to 29/10/2008 and includes data at weekly and monthly frequencies. This allows us to compare risk aversion and hedging scenarios for hedges held over different time periods to reflect the different holding periods of hedgers. All data were obtained from Commodity Systems and returns were calculated as the differenced logarithmic prices. A continuous series was formed with the contract being rolled over by largest volume. Descriptive statistics for each series are displayed in Table 1, while the pattern of the weekly Unleaded Gasoline price series can be seen from Fig 1[16].

**[TABLE 1 HERE]**

**[FIG. 1 HERE]**

On examining the general characteristics of the return distributions, we see a positive mean for each series indicating the strong price rises over the period. This can be attributed to a fall in the surplus production capacity of oil, which fell from 7 million barrels per day in January of 2002, to less than 1 million barrels per day in October of 2004 (Stevens, 2005). The most notable feature of the unleaded gasoline series is a large increase in both price and associated volatility during September 2005. This can be attributed to concerns about the supply of refined oil products, such as gasoline, which followed in the aftermath of hurricane Katrina. The period from September 2005 to June 2008 showed further strong increases. This period was characterised by further uncertainty about the supply of oil based products due to large increases in demand attributed to development in India and China. Speculation was also a key driver of the price and volatility during this period. The large changes in price and the rise in volatility is ideal from our perspective as it allows us to examine whether hedging would be effective in such circumstances (see Section 5). The lower frequency (monthly) data exhibits both higher means and variances, as compared with the weekly data, indicating that there may be differences in both the observed risk aversion, and hedge strategies for different hedging intervals. Also, each of the data frequencies is non-normal, with the monthly frequency in particular, displaying significant skewness. Therefore, we would expect to see differences in hedging strategies and outcomes for short as compared with long hedgers. LM tests with 4 lags were used to check for ARCH effects with significant ARCH effects present for the weekly series only.

---

[14] The contract used was the (HU) contract and not the reformulated blendstock (RB) which began trading in October 2005. The HU contract was still the dominant contract throughout the remainder of 2005 in terms of volume traded and therefore was continued throughout the sample period.
[15] The contract trades in units of 42,000 gallons on the NYMEX through open outcry. Further details of the contract and its trading characteristics are available at www.nymex.com/HU_spec.aspx

[16] Monthly data exhibit similar patterns.

**Estimation Procedure**

We estimate the CRRA by fitting the GARCH-M model to unleaded gasoline data. We then estimate risk preferences that are appropriate to the energy hedging context. To estimate the utility maximizing OHR for use in period $t$, from (2) we require estimates of $E(r_{ft})$, $\sigma_{ft}^2$ and $\sigma_{sft}$ as well as the estimated risk aversion parameter $\lambda$. In this paper we generate two different sets of OHR's. The first OHR for use in period $t$ is estimated using data available up to time $t$. To allow the hedge ratio to vary over time, we have adopted a rolling window approach using a window length of 10 years (with sufficient observations at a monthly frequency). We generate 174 1-period hedges for time t at the weekly frequency and 44 hedges at the monthly frequency for the period February 2002 to June 2005. The CRRA estimation is based on the hedged portfolio return. We use a hedge ratio of zero which effectively means that the short hedgers risk aversion is estimated using the underlying spot asset (unleaded gasoline). This is appropriate given that short hedgers have a long position in the unleaded gasoline market and this should provide an appropriate measure of their risk aversion. For a long hedger, they are long the unleaded gasoline futures contract, therefore their risk aversion is based on the unleaded gasoline futures. Using these assets, we can estimate the implied CRRA for short and long hedgers respectively, using (12). The sample is then rolled forward by one observation, keeping the window length unchanged. In this way, the OHR is continually updated by conditioning on recent information. We then obtain a time-varying utility maximizing OHR that incorporates observed risk aversion in the energy market – the RAHR. We also calculate a time-varying MVHR using the same rolling window methodology but employing (3). This doesn't incorporate the risk aversion parameter and is based on the variance covariance matrix alone.

The second set of OHR's we estimate are 1-step ahead forecast hedges for use in period t+1. We use time varying estimates of risk aversion to generate forecasts of future risk aversion that allows us to estimate utility maximizing hedge strategies today that reflect the future concerns of hedgers. To do this, we reserved a sub-period of three years of data at both the weekly and monthly frequencies to allow us to generate forecasted OHR's in a consistent manner. We used the estimates from the t-period hedges to generate hedges for use in period t+1. This procedure enabled us to generate a hedge ratio for use today and for tomorrow. The time-varying hedges were forecast with both the CRRA and the expected return on futures postulated to follow an AR (1) process, while a random walk process was assumed for the MVHR. Using this methodology, we obtained 174 hedges for use in period t+1 at the weekly frequency and 43 hedges at the monthly frequency. These covered the period from July 2005 to October 2008. The advantage of our approach is that it allows us to generate sufficiently large numbers of hedges for analysis using relatively low frequency data while examining the hedging strategies for long and short hedgers separately.

**Hedging Effectiveness**

We examine hedging effectiveness by forming hedged portfolios for both short and long hedgers using (1a and 1b) together with the OHR's as estimated from our models. The returns from these hedging portfolios are then used to examine performance, using a measure of hedging effectiveness based on the expected utility of the hedged returns. This is calculated as the difference between the expected utility of the hedged and un-hedged portfolios:

$$EU_{HedgedPortfolio} - EU_{UnhedgedPortfolio} \qquad (16a)$$

$$\text{EU} = E(R_{pt}) - 0.5(\lambda \sigma_{pt}^2) \tag{16b}$$

where $E(R_{pt})$ is the mean return on the portfolio, and $(\lambda \sigma_{pt}^2)$ is the risk aversion parameter multiplied by the variance of the portfolio (Sharpe, 1992). The performance measures are based on the expected utility performance criterion. This is derived using this mean CRRA value, as well as the mean and variance of the hedged portfolios.

## 5. Empirical Findings

In this section, we examine our findings for both short and long hedgers using both the weekly and monthly hedging frequencies. Firstly, we examine the estimated risk aversion. Secondly, we examine optimal hedge strategies and finally, we look at hedging effectiveness.

**Estimated Risk Aversion**

Our key element is the estimation of the time-varying coefficient of relative risk aversion based on the observed risk preferences of two different classes of hedgers. Results are presented for both short and long hedgers and for both weekly and monthly hedging intervals in Fig. 2.

[FIG. 2 HERE]

From Fig. 2, it is obvious that the observed CRRA is strongly positive, indicating that the relationship between volatility and expected return is positive.

We can also see that the observed risk aversion of both sets of hedgers tends to vary over time, which indicates that the amount of compensation that is required by risk averse investors for bearing risk is not constant. This tends to support the findings in the habit formation and asset pricing literatures on the time-varying nature of conditional risk aversion (see, for example, Campbell and Cochrane, 1999, Brandt and Wang, 2003). There is also an upward trend in risk aversion indicating that investors have become more risk averse over the period. Intuitively this is appealing as it indicates that investors in oil markets are reacting to their increased uncertainty in the wider global economy that has characterised the time period we examine. Also, differences in the observed risk aversion for each of the different hedging intervals reflect the fact that the time series properties of different frequency data tend to differ (Glosten et al, 1993). To further examine the dynamics of the CRRA estimates, summary statistics are presented in Table 2.

## [TABLE 2 HERE]

On examining the general characteristics of the risk aversion parameter estimates, we find that for the weekly frequency they range from an average of 0.34 to 0.40 for the short and long hedgers respectively. For the monthly hedges the mean CRRA's are higher at 0.43 for the short hedgers and 0.56 for the long hedgers, and are significantly different from each other. Therefore, hedgers with longer hedging time horizons tend to be more risk averse. This is not surprising given that the volatility in the energy market changes very quickly, and those who have shorter investment horizons tend to take on more risk than those who invest over longer periods. This may also indicate that a higher proportion of investors with shorter time horizons are speculators. From Table 2 we can also see that the CRRA of investors who hedge shorter intervals exhibit much greater dispersion, thus supporting the idea that they are perhaps more influenced by short term considerations. The findings also show that long hedgers exhibit greater risk aversion at both frequencies. This result may be related to the fact that short hedgers are producers of unleaded gasoline which is relatively price inelastic, whereas long hedgers as consumers are more exposed to price changes and may be more risk averse as a result. This result is consistent with Devlin and Titman (2004) who find evidence that consumers tend to be more active on the hedging side than producers. The differences in risk aversion for different types of hedger should yield different hedging strategies given that the hedging strategy is driven by concern for the opposite sides of the return distribution. This will be discussed in the next section.

We note that our mean CRRA estimates are lower than the risk aversion estimates that have been used for equities. This is because the hedging literature has relied on the estimates of the CRRA taken from either the consumption or asset pricing literatures, where the risk aversion is based on the equity risk premium, rather than being specifically calculated for a set of investors, such as hedgers who have an exposure to a single asset. Furthermore, many of the papers that have estimated risk aversion have used much shorter data sets whereas we use a 10 year window period. Therefore our approach will differ from hedging strategies that use risk aversion values such as those found in the general literature. In terms of a comparison, our estimates are probably closest to those of Brandt and Wang (2003) who estimated mean relative risk aversion parameters in the range 0.81 to 1.43 for monthly data[17].

---

[17] To investigate this issue further we also estimated risk aversion using S&P500 index data employing the same methodology that we used to obtain the risk aversion of energy hedgers over the same period. The results we obtained were similar to those used in the equity literature, with risk aversion values in the region of 2 – 5 (full results are available on request).Therefore we find that the risk aversion values of equities are different and should not be applied to energy products. A reviewer has pointed out that our analysis using single assets deviates from the approach in the equity literature where a market portfolio utilizes a highly diversified portfolio of assets. However the approach followed here has also been applied in settings with few assets (eg. for results on currency markets see Giovannini and Jorion, 1989).

**Optimal Hedging Strategies**

Fig 3 plots a comparison between the time-varying hedge strategies for both the RAHR and the MVHR for weekly and monthly hedges. Summary statistics on the different hedge ratios are presented in Table 3. From Fig 3, it is immediately apparent that the time-varying nature of the CRRA has influenced the RAHR with weekly hedges showing the largest variation. This is further emphasised when we examine the range of the RAHR's in Table 3. For example, the short hedgers at the weekly frequency, the OHR ranges from -0.082 to 0.561 with a mean value of 0.296. This means that for each unit of the spot asset held, the short hedger sold an average of 0.296 futures contracts.

[FIG. 3 HERE]

[TABLE 3 HERE]

This compares with a range of just -0.244 to 0.685 (mean value 0.244) for the monthly hedges. This large spread is indicative of the impact that lower values of risk aversion can have on the optimal hedge as the speculative element can cause the ratio to be quite volatile and in some cases such as for short hedgers, this results in hedgers leveraging up their exposure, for example when the hedge ratio is below zero.

For long hedgers the effect of the speculative element of the hedge ratio is quite pronounced and results in hedges well in excess of 1. For example, the average RAHR is 1.62 and 1.50 at the weekly and monthly frequencies respectively. In effect this means that taking the expected return on futures into account causes the long hedgers to exchange their long futures exposure to long spot exposure. This result is consistent with deVille deGoyet, Dhaene and Sercu, (2008) who estimate utility maximising hedge ratios for commodities ranging from 1.68 to 2.37. We also observe large differences in the hedge strategies of the risk minimising hedgers (MVHR) and the utility maximising hedgers (RAHR). For short hedgers the RAHR is below the MVHR whereas for long hedgers the RAHR is above the MVHR. Again, this indicates that the speculative component of the hedge ratio plays a considerable part in determining the optimal hedge strategy when risk aversion is taken into account.

Table 3 also presents a statistical comparison of the different hedges. Firstly, we compare the RAHR with the MVHR within each set of hedgers to see whether incorporating risk aversion makes a significant difference to the optimal hedge strategy. The results indicate that there are significant differences between the RAHR and the MVHR for each hedging interval, and for both short and long hedgers. The differences are most pronounced for weekly hedges but even at the monthly frequency the differences are still significant. In economic terms the differences are also significant. For example, using the weekly hedging frequency for a short hedger, the mean hedge ratio adopted by a utility maximizing hedger is 0.296, as compared with 1.006 for the risk minimising hedger. Thus the two strategies are completely different in terms of the number of futures contracts that will be used to hedge a spot position. This result indicates that when explicit risk aversion is incorporated into the calculation of the optimal hedge ratio, there will be large differences between the utility maximizing and risk minimising strategies.

This finding contrasts with the finding in Chen et al (2001), who find little difference between the utility maximizing and risk minimising OHR's, and emphasises the importance of using explicit risk aversion estimates rather than relying on values drawn arbitrarily from other literatures. Indeed, it indicates that hedgers who ignore the part played by risk aversion in determining the optimal hedge ratio may choose hedge strategies that are suboptimal in terms of expected utility. On the basis of this evidence, it would seem that much larger values of the CRRA than those that we have observed would be required for the RAHR to converge towards the MVHR. This is clearly in evidence if we examine Fig 4 which plots the relationship between the CRRA and the RAHR for short hedgers. Here is evidence of a positive relationship implying that as risk aversion increases, the RAHR will converge towards the MVHR.

[FIG. 4 HERE]

Secondly, we compare the mean RAHR for short hedgers as compared with long hedgers. The results indicate that there are significant differences between the RAHR of short and long hedgers for both weekly and monthly hedging frequencies. The RAHR is lower for short hedgers in all cases. We also examine the relationship between risk aversion and volatility in Fig. 4. There is a positive relationship between the time-varying volatility and the CRRA indicating that as volatility increases, risk aversion increases. This finding is intuitively appealing is it indicates that hedgers become more risk averse during times of high volatility which in turn causes them to hedge more as the risk minimising element of the hedge strategy comes to dominate the speculative element.

[INSERT FIG. 5]

**Hedging Performance**

To compare the hedging effectiveness of both the RAHR and the MVHR, we use the difference in the expected utility of the hedged strategies, as compared with an un-hedged strategy. Note that this measure of hedging effectiveness is dependent on the risk aversion parameter applied and since we are using a time-varying method we had to consider which of the risk aversion values to use. We have decided to use the mean CRRA as this best represents the average expected utility.

**[TABLE 4 HERE]**

Examining first the in-sample results for short hedgers, from Table 4 we can see that the RAHR dominates the MVHR in terms of expected utility. Also, in all cases the expected return for the RAHR is greater. For example, at the weekly frequency, the mean return for the RAHR is 0.12% as compared with -0.04% for the MVHR, however, when the performance criterion is risk minimisation alone, the MVHR dominates. For long hedgers the results are reversed with the MVHR dominating the RAHR in terms of expected utility. When we compare the risk aversion hedge strategies of short Vs long hedgers in terms of hedging effectiveness, we find that the results favour the long hedgers for the weekly and monthly hedges. In each case both the RAHR and MVHR perform better for long as compared with short hedgers. These results support the findings in Demirer et al (2005) which indicate that long hedgers outperform short hedgers.

We now focus on the out-of-sample or forecast performance of the hedging models. From Table 4, comparing the MVHR's and the RAHR's in terms of expected utility, the MVHR outperforms the RAHR in all cases. These results support the findings in Brooks Cerny and Miffre (2007) who find that hedges that incorporate risk aversion fail to consistently outperform OLS in an out-of-sample setting. If risk reduction alone were the hedging effectiveness criterion, the MVHR would be the clear winner overall. It yields the lowest risk of each of the hedging strategies in every single case. This result mirrors the findings in the general hedging literature where hedging effectiveness is based on minimising risk. Finally, when we compare the out-of-sample risk aversion hedge strategies of short vs. long hedgers, we find that short hedgers do better in terms of hedging effectiveness than long hedgers. For example from column 1, the hedging effectiveness of a short hedger for the RAHR is 0.20% as compared with -0.09% for a long hedger. This result holds for all hedges and for both weekly and monthly frequencies.

## 6. Conclusion

In this paper we put forward a method for calculating and applying the observed risk aversion of energy hedgers in formulating a hedging strategy. Our focus on energy hedging is timely, given the importance of the market for energy and the increasing uncertainty surrounding energy prices going forward. The risk aversion parameter is a key input into the utility maximizing hedging framework. However, despite its importance it has been applied in the hedging literature in an arbitrary manner. We estimate a time-varying coefficient of relative risk aversion, based on the observed risk preferences of both short and long hedgers. This allows us to calculate and apply OHR's that reflect the risk preferences of hedgers.

Our most important finding is that there are significant differences between the RAHR and the MVHR both statistically and economically. This means that when explicit risk aversion is taken into consideration, there will be large differences in the expected utility and risk minimising hedge strategies. In terms of risk aversion, the mean CRRA estimates of hedgers were generally lower than the values cited in other literatures (e.g. equity) on risk aversion and this gave rise to the large differences between the MVHR and the RAHR.

Differences also emerged in terms of the risk preferences of short as compared with long hedgers. In general, long hedgers are more risk averse than short hedgers. This finding is intuitively appealing as it supports the view that commodity users tend to hedge more than producers. In addition, we found differences in the risk aversion, and therefore the hedging strategies depending on the hedging interval with higher risk aversion being exhibited by investors with longer time horizons. Furthermore, when the observed risk aversion parameters were used to calculate OHR's, we found that long hedgers tended to outperform short hedgers in terms of expected utility at both weekly and monthly frequencies in an in-sample setting. In terms of overall performance in-sample, the RAHR tends to dominate the MVHR in terms of expected utility for short hedgers only. Out-of-sample, the MVHR tends to outperform the RAHR.

These findings indicate that the level of risk aversion is too important to be chosen arbitrarily as it will have a pronounced effect on the choice of energy hedging strategy. Risk aversion values that are applied in a hedging context should not be taken from the asset pricing literature, but should be estimated based on the risk preferences of energy hedgers themselves. They also indicate the importance of tailoring hedge strategies to take account of the length of the hedge as well as the type of hedger.

This paper has examined the impact of risk aversion on the hedging decision by illustrating a procedure for obtaining a time varying risk aversion based hedge ratio for the quadratic utility framework. We also recognise that there are many other utility functions that could also yield useful insights. An examination of the relationship between risk aversion and hedging for other utility functions provides a possible avenue for further work in this area.

|         | Mean   | Stdev | Min    | Max   | Skewness | Kurtosis | B-J   | LM    |
|---------|--------|-------|--------|-------|----------|----------|-------|-------|
| **WEEKLY** |        |       |        |       |          |          |       |       |
| SPOT    | 0.0012 | 0.059 | -0.215 | 0.206 | -0.108   | 0.506*   | 11.0* | 28.5* |
| FUTURES | 0.0011 | 0.050 | -0.187 | 0.161 | -0.091   | 0.558*   | 12.5* | 17.8* |
| **MONTHLY** |      |       |        |       |          |          |       |       |
| SPOT    | 0.0047 | 0.118 | -0.329 | 0.301 | -0.367*  | 0.215    | 5.32  | 3.8   |
| FUTURES | 0.0043 | 0.103 | -0.366 | 0.284 | -0.548*  | 1.062*   | 21.2* | 3.2   |

**Table 1: Summary Statistics for Unleaded Gasoline Spot and Futures Returns**
Summary statistics are presented for the log returns of each spot and futures series. The total sample period runs from 09/02/1992 until 05/11/2008. The Bera-Jarque (B-J) statistic combines skewness and kurtosis to measure normality. LM, (with 4 lags) is the Engle (1982) ARCH test. The test statistic for B-J and LM tests are distributed $\chi^2$. * denotes significance at the 1% level.

|  | CRRA | |
|---|---|---|
|  | SHORT HEDGERS | LONG HEDGERS |
| **WEEKLY** | | |
| MEAN | 0.339*† | 0.396† |
| MIN | 0.100 | 0.107 |
| MAX | 0.650 | 0.721 |
| STDEV | 0.137 | 0.168 |
| **MONTHLY** | | |
| MEAN | 0.425* | 0.561 |
| MIN | 0.135 | 0.174 |
| MAX | 0.751 | 0.879 |
| STDEV | 0.179 | 0.230 |

**Table 2:    Risk Aversion of Short and Long Hedgers**

CRRA is the estimated risk aversion parameter, summary statistics are presented for the in sample period. Statistical comparisons are drawn between the Mean CRRA value for short and long hedgers across each hedging interval. There are significant differences between the CRRA values of short Vs long hedgers at both weekly and monthly frequencies. Also the CRRA differs significantly between frequencies. * denotes significance at the 1% level for a comparison of the risk aversion of short Vs long hedgers. † denotes significance at the 1% level for comparison of the CRRA for weekly Vs monthly frequencies.

|  | Panel A: Short Hedgers | | Panel B: Long Hedgers | |
| --- | --- | --- | --- | --- |
|  | (1) | (2) | (3) | (4) |
|  | RAHR | MVHR | RAHR | MVHR |
| ROLLING WINDOW HEDGES | | | | |
| **WEEKLY** | | | | |
| MEAN | 0.296†* | 1.006 | 1.62† | 1.006 |
| MIN | -0.082 | 0.989 | 1.38 | 0.989 |
| MAX | 0.561 | 1.018 | 2.37 | 1.018 |
| STDEV | 0.112 | 0.007 | 0.145 | 0.007 |
| **MONTHLY** | | | | |
| MEAN | 0.516†* | 1.077 | 1.50† | 1.077 |
| MIN | 0.244 | 1.052 | 1.37 | 1.052 |
| MAX | 0.685 | 1.122 | 2.01 | 1.122 |
| STDEV | 0.111 | 0.019 | 0.113 | 0.019 |

**Table 3: Risk Aversion and Hedge Strategies of Short and Long Hedgers**

RAHR is the risk aversion hedge ratio; MVHR is the minimum variance hedge ratio. Summary statistics are presented for the in sample period. Two statistical comparisons are drawn. We first compare the mean hedge ratios of short Vs long hedgers. Using the Weekly interval for example, there is a significant difference between the short hedgers mean hedge ratio and the long hedgers hedge ratio. We also compare the RAHR and MVHR within each set of hedgers. Using the monthly interval for example, in columns 1 and 2, the RAHR is significantly different from the MVHR. * denotes significance at the 1% level respectively for short Vs long comparison. † denotes significance at the 1% level for comparison of the RAHR and MVHR.

|  | Panel A: Short Hedgers | | | Panel B: Long Hedgers | | |
| --- | --- | --- | --- | --- | --- | --- |
|  | (1) HE (x10$^{-2}$) | (2) HE (x10$^{-2}$) | (3) HE (x10$^{-2}$) | (4) HE (x10$^{-2}$) | (5) HE (x10$^{-2}$) | (6) HE (x10$^{-2}$) |
|  | RAHR | MVHR | NO HEDGE | RAHR | MVHR | NO HEDGE |
|  | | | IN-SAMPLE | | | |
| **WEEKLY** | | | | | | |
| MEAN | 0.12 | 0.04 | 0.55 | -0.14 | -0.04 | -0.55 |
| SD | 4.34 | 2.48 | 5.52 | 4.36 | 2.48 | 5.52 |
| EU | 0.09 | 0.03 | 0.50 | -0.17 | -0.06 | -0.61 |
| HE | -0.41 | -0.47 | 0.00 | 0.44 | 0.56 | 0.00 |
| **MONTHLY** | | | | | | |
| MEAN | 0.14 | 0.02 | 2.00 | -0.12 | -0.02 | -2.00 |
| SD | 6.47 | 4.46 | 10.39 | 8.05 | 4.46 | 10.39 |
| EU | 0.06 | -0.02 | 1.78 | -0.30 | -0.08 | -2.30 |
| HE | -1.72 | -1.80 | 0.00 | 2.00 | 2.23 | 0.00 |
|  | | | OUT-OF-SAMPLE | | | |
| **WEEKLY** | | | | | | |
| MEAN | 0.08 | 0.06 | -0.07 | -0.08 | -0.06 | 0.07 |
| SD | 4.00 | 2.50 | 6.12 | 4.46 | 2.50 | 6.12 |
| EU | 0.04 | 0.04 | -0.17 | -0.16 | -0.08 | -0.07 |
| HE | 0.20 | 0.21 | 0.00 | -0.09 | -0.01 | 0.00 |
| **MONTHLY** | | | | | | |
| MEAN | -0.19 | 0.34 | -0.05 | -0.58 | -0.34 | 0.05 |
| SD | 8.27 | 3.66 | 12.27 | 7.62 | 3.66 | 12.27 |
| EU | -0.41 | 0.30 | -0.55 | -0.81 | -0.40 | -0.55 |
| HE | 0.49 | 1.44 | 0.00 | -0.06 | 0.45 | 0.00 |

**Table 4:	Hedged Returns and Hedging Performance – Time Varying Hedges**

Mean, Standard Deviation (SD), Expected utility (EU) and Hedging Effectiveness (HE) are presented for the RAHR, the MVHR and a No Hedge position. HE reflects the difference in the EU of both the RAHR and MVHR as compared with the No hedge position. A positive value indicates an effective hedging outcome as compared with a No Hedge position.

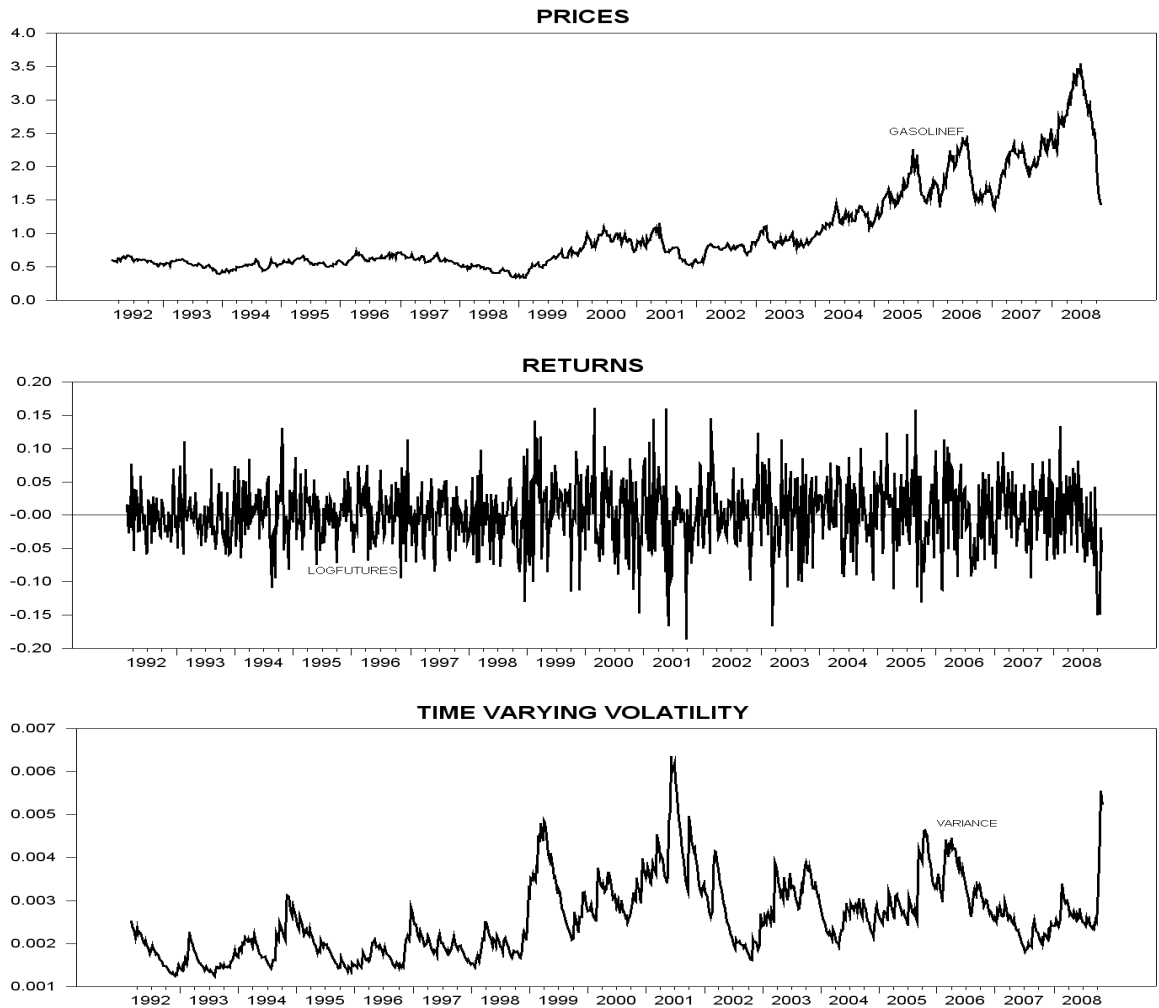

**Fig 1: General Data Characteristics**

Fig 1 displays the general data characteristics for the weekly unleaded gasoline spot series. Each series is shown for the period from 06/03/2002 to 29/10/2008. Volatility is obtained from fitting a GARCH (1, 1) model. Note the strong increase in price for the period 2002 - 2008 and the associated increase in volatility.

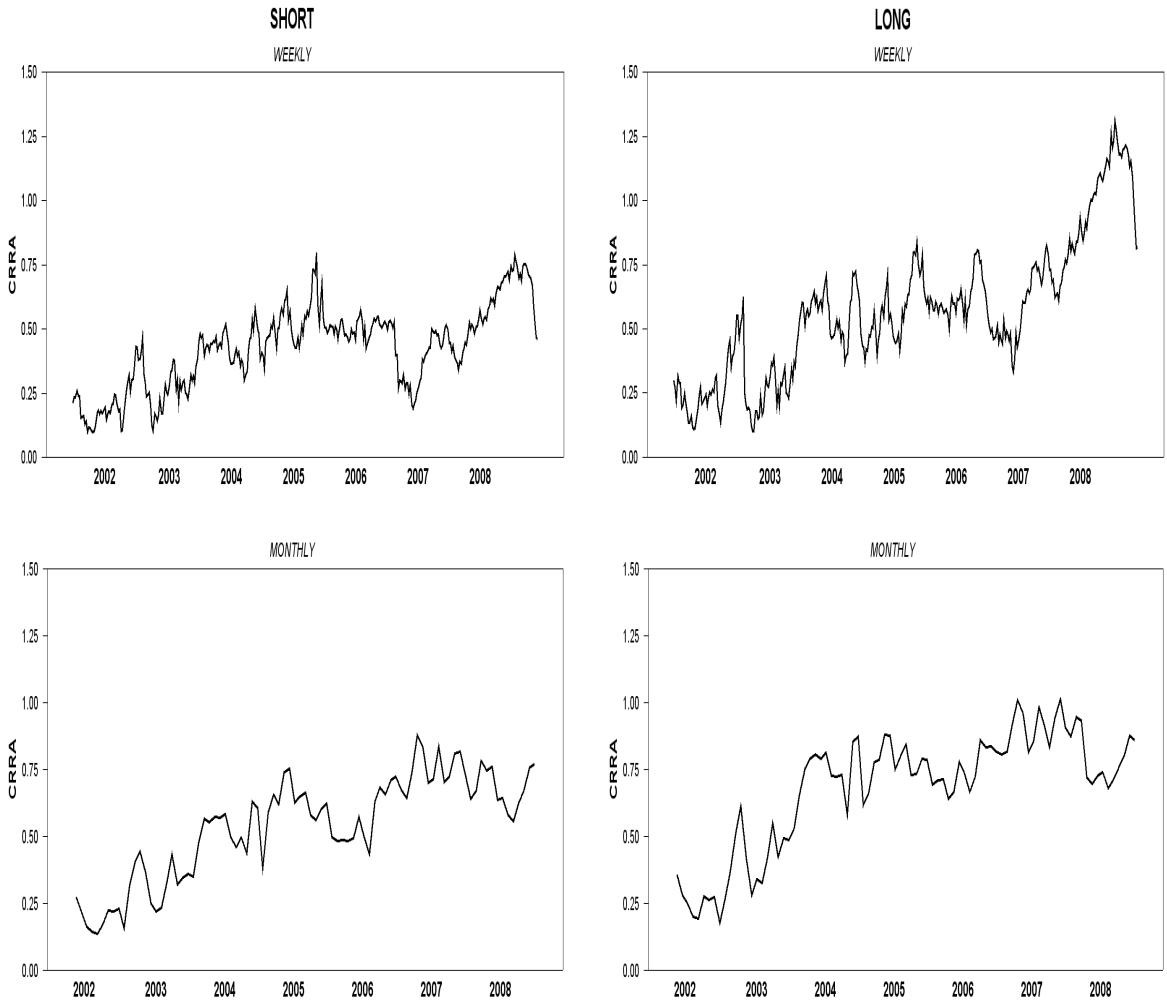

**Fig 2: Time-varying Coefficient of Relative Risk Aversion for the Unleaded Gasoline contract**

The CRRA is plotted for both short and long hedgers and for the weekly and monthly hedging intervals. The risk aversion of short hedgers is based on their long position in the spot asset whereas the risk aversion of the long hedgers is based on their long position in the futures contract.

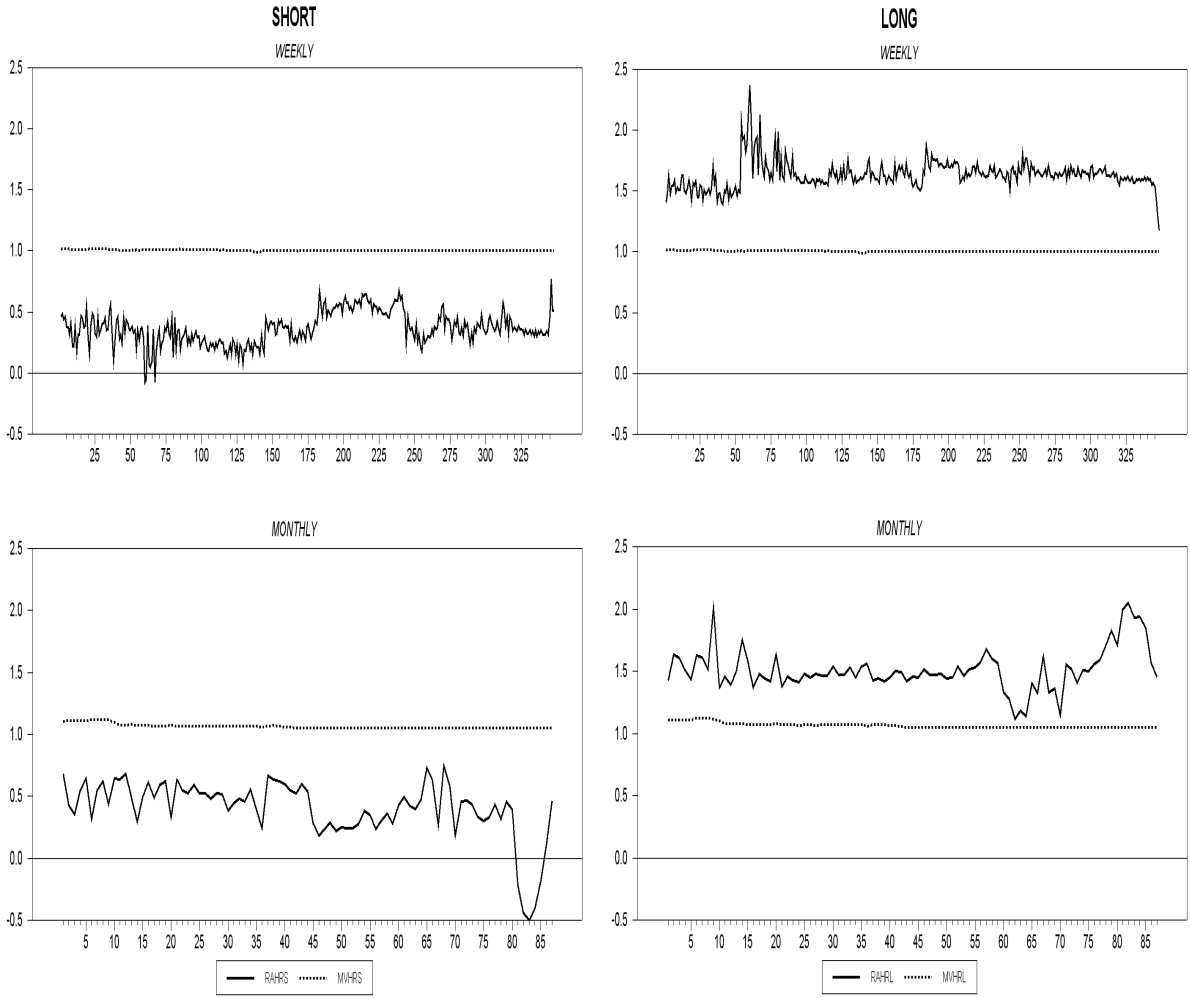

**Fig 3: Time-varying Optimal Hedge Ratios for the Unleaded Gasoline contract**

This figure plots both the time varying Risk Aversion Hedge (RAHR) and the time varying minimum variance hedge (MVHR) for both short and long hedgers for both weekly and monthly hedging frequencies.